\documentclass[11pt]{article}

\usepackage[margin=1in]{geometry}
\usepackage{lineno}
\usepackage{bm}
\usepackage{amsmath}
\usepackage{amssymb}
\usepackage{graphicx}
\usepackage{authblk}
\usepackage{hyperref}

\graphicspath{{figures/}}

\title{Coincidence-based spectral engineering for spectral matching in cascaded downconversion}

\author[1]{Mohammed Charafi}
\author[1]{Alexandre Z. Leger}
\author[1,2]{Samridhi Gambhir}
\author[1,*]{Deny R. Hamel}

\affil[1]{D\'epartement de physique et d'astronomie, Universit\'e de Moncton, Moncton, New Brunswick E1A 3E9, Canada}
\affil[2]{Quantum Bridge Technologies Inc., Toronto, Ontario M5G 0C6, Canada}
\affil[*]{Corresponding author: \texttt{deny.hamel@umoncton.ca}}

\date{}

\begin{document}

\maketitle

\begin{abstract}
Three-photon states generated via cascaded spontaneous parametric downconversion provide a direct route to multipartite entanglement. However, current implementations require careful spectral matching between successive nonlinear stages, which constrains the choice of downconversion sources. In this work, we show that coincidence-based spectral filtering relaxes this requirement by conditionally tailoring the spectrum of the pump photon entering the second stage. By filtering the herald photon, we conditionally tailor the spectrum of its partner to match the acceptance bandwidth of the secondary nonlinear process, enabling efficient coupling between broadband and narrowband stages without altering the sources themselves. Using an electro-optically gated spectrometer, we directly measure the conditional spectra that govern the cascaded process, allowing us to quantitatively predict the enhancement in second-stage conversion probability per detected herald. We then verify this prediction through photon-triplet measurements, demonstrating improved performance at fixed heralding rates. Our results establish coincidence-based spectral engineering as a practical tool for optimizing cascaded downconversion, particularly in regimes limited by detector saturation or spectral incompatibility.
\end{abstract}

\section{Introduction}
The generation of correlated multi-photon states is a central capability for photonic quantum technologies, enabling applications such as quantum communication, quantum metrology, and photonic quantum information processing~\cite{pirandolaAdvancesQuantumCryptography2020,barbieriOpticalQuantumMetrology2022,genoveseRealApplicationsQuantum2016,flaminiPhotonicQuantumInformation2018}.
In particular, three-photon states provide access to multipartite entanglement~\cite{bouwmeesterObservationThreephotonGreenbergerHorneZeilinger1999,eiblExperimentalRealizationThreeQubit2004} and enable protocols such as quantum secret sharing~\cite{Hillery1998,Bell2014},  heralded Bell-state generation~\cite{hamelDirectGenerationThreephoton2014} and tests of nonlocality~\cite{Erven2013,panMultiphotonEntanglementInterferometry2012}.
Cascaded spontaneous parametric downconversion (C-SPDC) provides a direct approach to producing such states, by using a photon generated in a first SPDC process as the pump for a second nonlinear interaction~\cite{greenbergerBellTheoremInequalities1990,hubelDirectGenerationPhoton2010}, as shown in Fig.~\ref{fig:Fitltre mode0-1} (a).
This method has been shown to produce and transform entangled states with high fidelity across multiple degrees of freedom, including in polarization~\cite{chaissonPhasestableSourceHighquality2022}, energy-time~\cite{agne2017observation}, and orbital angular momentum~\cite{Kopf2025}.
It can also serve as an enabling resource for photon precertification, a promising tool for overcoming transmission losses in photonic quantum networks~\cite{cabelloLoopholeFreeBellTest2012,meyer-scottCertifyingPresencePhotonic2016}.
In all practical implementations, however, the overall triplets rates produced by C-SPDC have remained extremely low, motivating efforts to improve the compatibility and performance of cascaded photon-pair sources~\cite{Leger2023}.

Experimental realizations of C-SPDC have so far combined various types of sources; the process has been demonstrated on single platforms with both nonlinear processes in periodically poled lithium niobate (PPLN)~\cite{krapickOnchipGenerationPhotontriplet2016,Harper2026}, while other experiments have  used PPLN only for the second stage and a different nonlinear medium for the first stage, including a rubidium vapour~\cite{dingHybridcascadedGenerationTripartite2015} or periodically poled potassium titanyl phosphate (PPKTP)~\cite{hubelDirectGenerationPhoton2010}. In these various implementations, the choice of conversion media was limited by the need to coordinate the spectral properties of these independent sources; any mismatch between the bandwidth of produced photons in the first source and the acceptance bandwidth of the secondary nonlinear medium can lead to a significant fraction of photons from the first downconversion being incompatible with the second stage. For example, Hubel et al. estimated that this effect~\cite{hubelDirectGenerationPhoton2010} reduced detected triplet rates by 35\% reduction, despite choosing sources with relatively well matched spectral properties.

Although the reduction in rates in this case can appear modest, it is poised to become more problematic as steps are taken to improve the overall process efficiency of C-SPDC.
For example, replacing the first stage with a high-brightness source \cite{Pollmann26} could increase the \textit{generated} triplet rate by orders of magnitude, as higher spectral brightness would provide more photon pairs within the acceptance bandwidth of the second stage. However, such sources typically have a much broader emission spectrum, so most herald detections would correspond to partner photons that lie outside the second-stage acceptance bandwidth and therefore would not contribute to the cascade. As a result, the heralding detector would become saturated by these uncorrelated heralds long before the corresponding increase in the \textit{detected} triplet rate would be realized.

Alternatively, the secondary stage's nominal efficiency could be enhanced by increasing the secondary stage's  crystal length, but this would also reduce its acceptance bandwidth, thus limiting efficiency gains. 
Similarly, the secondary crystal's conversion efficiency could also be increased by  placing it inside a cavity, with an even more drastic effect on acceptance bandwidth~\cite{slatteryBackgroundReviewCavityEnhanced2019,Moqanaki2019,Puvsavec2025}.
These examples illustrate that spectral matching is currently a major design constraint in C-SPDC. The first and second nonlinear stages cannot generally be optimized independently, since improving one stage often changes its bandwidth in a way that reduces compatibility with the other. Source selection is therefore dictated in part by the need for closely matched spectral bandwidths, at the expense of optimizing brightness or conversion efficiency.

The present work aims to remove this design constraint by making spectral compatibility an experimentally adjustable parameter rather than an inherent property of the selected sources.
To achieve this, we implement an approach to match broadband primary sources with secondary stages that have a narrow acceptance bandwidth by using spectral filtering of photon pairs in coincidence, and demonstrate its ability to improve the rates of detected triplet from C-SPDC.
By filtering in the heralding arm, the number of detected heralds whose partners lie outside the second-stage acceptance is reduced. In an experiment limited by herald detector saturation, this filtering allows the first-stage pump power to be increased while maintaining a fixed heralding rate. A larger fraction of detected heralds thus correspond to photons capable of driving the second stage, leading to a higher detected triplet rate. 

To validate the method, we first measure the relevant coincidental spectra using a specially built optically gated spectrometer, allowing us to predict the expected triplet rate enhancement.
We then verify that this coincident filtering leads to higher photon-triplet generation rates.

\section{Theory}

At first glance, spectral filtering may not appear to offer an advantage for C-SPDC, as photons produced by the first SPDC stage that fall outside the acceptance bandwidth of the second nonlinear crystal will simply fail to convert and are effectively rejected by the second stage. In this sense, the second conversion already acts as a spectral filter on the photons entering the cascade. Removing these photons beforehand therefore seems unnecessary, since any benefit provided by a broadband source should already be present without additional filtering.

\begin{figure}[htpb!]
\centering\includegraphics[width=0.7\linewidth]{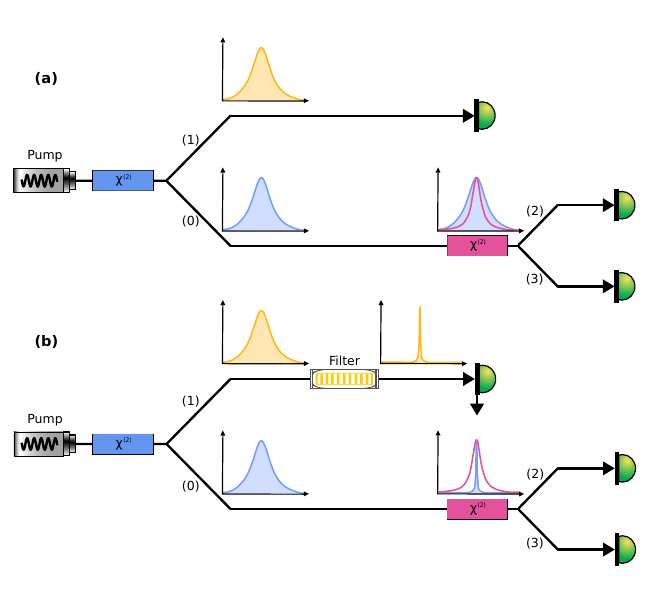}
\caption{C-SPDC with (a) no spectral filtering and (b) filtering mode in (1). Filtering in mode 0 would be ineffective because such a filter would simply replicate the spectral filtering effect of the nonlinear crystal. By using a narrowband pump, filtering in mode 1 produces a narrowing of the coincidence spectrum in mode, which can be aligned with the conversion peak of the second nonlinear crystal,  while the detection rate in mode 1 is reduced. 
}
\label{fig:Fitltre mode0-1}
\end{figure}

In practice, however, the detector used to herald the cascade has a maximum detection rate~\cite{hadfield2009,eisamanInvitedReviewArticle2011,Honjo2026}. When the heralding rate reaches this limit, detections whose partner photons fall outside the second-stage acceptance use detector capacity without contributing to triplet generation~\cite{hubelDirectGenerationPhoton2010,Kopf2025}.
This is illustrated in the configuration shown in Fig.~\ref{fig:Fitltre mode0-1}(a), where photons detected in mode 1 herald the presence of photons in mode 0 that are used to pump the second nonlinear crystal. When the primary SPDC source is broadband, many of these heralding detections correspond to photons in mode 0 whose wavelengths fall outside high-efficiency conversion region of the second crystal.
These detections therefore do not contribute to the generation of photon triplets, while still contributing to detector saturation.
Filtering the photons in mode 0, before they enter the second nonlinear crystal, does not resolve this issue; indeed, such filtering only reproduces the spectral selection already imposed by the phase-matching conditions of the second conversion and, therefore, does not improve detected triplet rates.

Instead, an advantage can be obtained by filtering the partner photon in mode 1 to enable conditional spectral shaping of the photon in mode 0~\cite{Meyer-Scott2017,Averchenko2020,Smith2026}. Similar spectral conditioning was recently used in the context of interfacing two SPDC sources~\cite{Harper2026}, although that work was limited to inferring cascaded generation indirectly through pair measurements rather than direct triplet detection. The idea, shown in Fig.~\ref{fig:Fitltre mode0-1}(b),  is to exploit the spectral correlations of photon pairs produced in the first SPDC process, so that selecting a narrow spectral region for photon 1 modifies the conditional spectrum of photon 0 observed in coincidence with that detection. If the pump laser is sufficiently narrowband the frequencies of the two photons are approximately constrained by energy conservation,

\begin{equation}
\omega_p \approx \omega_0 + \omega_1 ,
\end{equation}

\noindent where $\omega_p$ is the pump frequency. Then, if the joint spectrum of the bipohoton is sufficiently correlated, selecting photons in mode 1 within a spectral window restricts the possible frequencies of the corresponding photons in mode 0, such that the filter preferentially selects photons in mode 0 that are likely to lead to a downconversion in the second stage.

More formally, the spectral properties of photon pairs produced by the first SPDC process are governed by a joint spectral intensity $|f(\omega_0,\omega_1)|^2$. When photon 1 is detected through a filter with transmission function $T(\omega_1)$, the effective spectrum of photons in mode 0 conditioned on that detection becomes

\begin{equation}
S_0(\omega_0) \propto \int |f(\omega_0,\omega_1)|^2 T(\omega_1)\, d\omega_1 .
\end{equation}

The probability that a photon from the first stage successfully pumps the second conversion is  proportional to the overlap

\begin{equation}
P \propto \int S_0(\omega_0) A(\omega_0)\, d\omega_0 ,
\end{equation}

\noindent where $A(\omega_0)$ is the spectral acceptance function of the second nonlinear medium. The exepcted triplet rate is therefore given by 

\begin{equation}
R \propto R_\mathrm{h}\eta_2\eta_3\int S_0(\omega_0) A(\omega_0)\, d\omega_0 ,
\end{equation}
where $R_\mathrm{1}$ is the heralding rate in mode 1 and $\eta_i$ are the Klyshko efficiency in modes 2 and 3. By narrowing the conditional spectrum $S_0(\omega_0)$ so that it better overlaps the peak of $A(\omega_0)$, filtering in the heralding arm increases the probability that each detected photon contributes to a successful cascaded conversion. As a result, even though the filtering reduces the total photon flux, the rate of detected photon triplets can increase when the detection rate in mode 1 is held constant.

\section{Experimental demonstration} 

\begin{figure}[ht!]
\centering\includegraphics[width=0.75\textwidth]{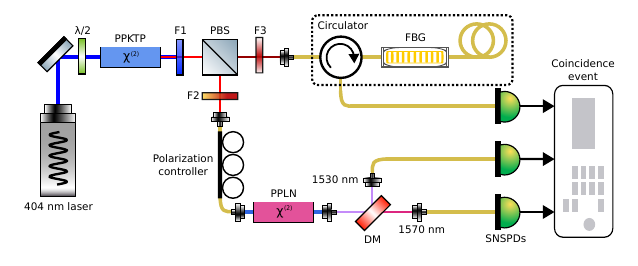}
\caption{Experimental setup for the measurement of photon triplets. The dashed box shows the optional spectral filtering of the heralding photon.
 }
\label{fig:Montage-triplets}
\end{figure}

Our experiment centers on quantifying the effect of inserting an appropriate filter in mode 1 of a C-SPDC setup, thus producing a narrower coincidence spectrum in mode 0, which should be more compatible with the second photon source.
As shown in Fig.~\ref{fig:Montage-triplets}, this C-SPDC setup starts with a narrowband 404~nm continuous wave laser (Toptica Topmode) pumping a 25~mm PPKTP crystal (Raicol crystals), producing photon pairs at 844~nm and 775~nm that are separated using a polarizing beam-splitter (PBS).
A colored- glass filter (F1, Thorlabs FGL715) and two bandpass filters (F2 and F3, Semrock FF01-769/41 and SemrockFF01-832/37), are used to spectrally select only the downconverted photons.
The 844~nm photons are then either filtered or detected directly, while the 775~nm photons are sent into a 50 mm PPLN reverse proton exchange waveguide (HC Photonics) for the second downconversion, producing photons at around 1530~nm and 1570~nm. 
At the output of this waveguide, the generated pairs are returned to free space and directed to a dichroic mirror (DM, Semrock BLP01-1550R) to separate the telecom photons.
For triplet measurements, all photons are detected using superconducting nanowire single-photon detectors (SNSPDs, Photon Spot), yielding combined coupling and detection efficiencies of 32\% for the $844$~nm photons and 30\% for the two telecom outputs. 
Triplet events are identified using timetags from a coincidence logic (Universal Quantum Devices), using a coincidence window of 1.875 ns.

Spectral filtering is implemented using a custom fiber Bragg grating (FBG) with a 100~$\pm$~25~pm  nominal bandwidth (AOS GmbH), chosen to be narrower than the expected acceptance bandwidth of the second stage. The filter is then combined with a fiber circulator (OZ Optics), to keep only the portion of the spectrum reflected by the FBG. 

\subsection{Characterization of filtered spectra}
\begin{figure}[htpb]
\centering\includegraphics[width=0.8\textwidth]{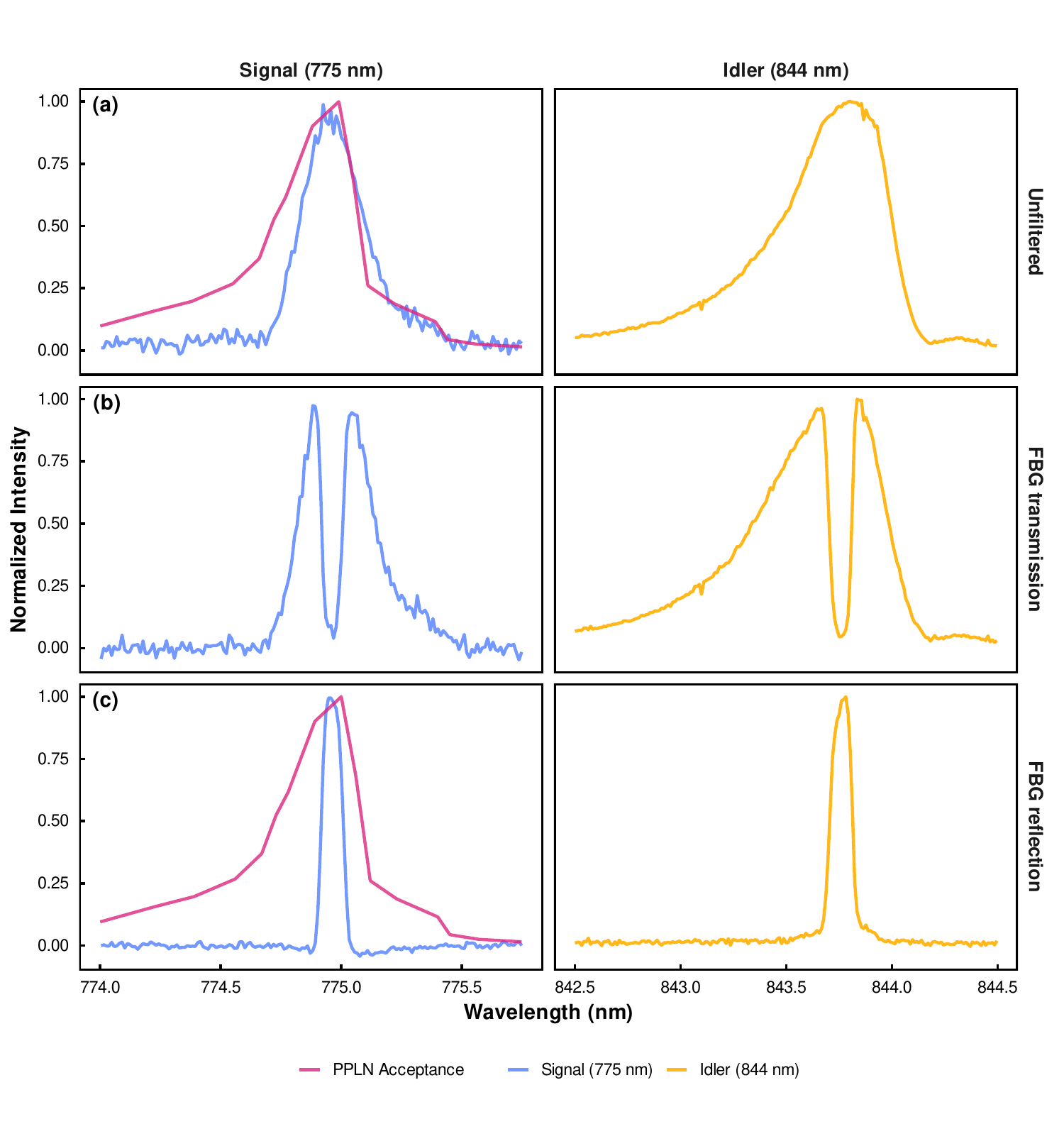}
\caption{Spectra at 844~nm and corresponding coincidence spectra at 775~nm. (a) Unfiltered spectra: 844~nm photons are directly sent to the spectrometer after being produced by the PPKTP. (b) Spectra transmitted through the FBG: 844~nm photons pass through the FBG before reaching the spectrometer. (c) Spectra reflected by the FBG: 844~nm photons reflected by the FBG are collected by the circulator before being detected by the spectrometer. For the 775~nm measurements, noise is reduced using background subtraction, where the background signal is obtained by adding additional delay before the Pockels cell is activated. 
}   
\label{fig:Spectres}
\end{figure}

To verify the expected gain in triplet rate from this approach, we first measure the unfiltered spectrum of 775~nm photons produced by the PPKTP crystal using a conventional grating spectrometer (LightMachinery). We compare this to the acceptance bandwidth of the PPLN waveguide, which is measured using a tunable narrowband laser (Toptica DL PRO 780), as shown in Fig.\ref{fig:Spectres}~(a). Based on these results, we calculate that in the unfiltered configuration the effective conversion efficiency is reduced by a factor of 0.61. 

Next, to characterize the effect of coincidence filtering, we measure the filtered spectrum at 844~nm in mode 1, as well as the conditional spectrum  at 775~nm remaining in mode 0 after post-selection. The latter measurement is not straightforward, as a conventional spectrometer placed in mode 0 would measure the full marginal spectrum of the photons produced by the first SPDC source, independently of whether their partners in mode 1 are detected within the selected spectral window. Accessing this spectrum therefore requires an experimental technique capable of selecting only those photons that occur in coincidence with the filtered detections.

\begin{figure}[htpb!]
\centering\includegraphics[width=0.75\textwidth]{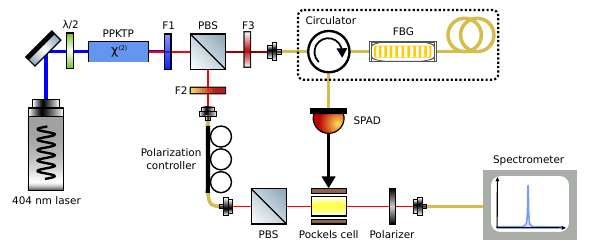}
\caption{Experimental setup for the coincidence spectrum measurement. After 844~nm photons have been filtered by the FBG and circulator, they are detected on a Si single-photon avalanche diode (SPAD). This detector is used to trigger a Pockels cell, which changes the polarization of the 775~nm photons from horizontal to vertical, allowing them to pass through the polarizer placed after the Pockels cell. The beam is then directed into optical fibers towards the spectrometer.
}
\label{fig:montage1}
\end{figure}

To achieve this, we use an electro-optic gating technique based on a Pockels cell~\cite{Leger2023}, as illustrated in Fig.~\ref{fig:montage1}. Photons at 844~nm are spectrally filtered and detected, and their detection signal is used to trigger the Pockels cell. When activated, the Pockels cell temporarily rotates the polarization of photons at 775~nm, allowing them to pass through a polarizer and reach the spectrometer. In this way, only photons in mode 0 that are temporally correlated with the filtered detections in mode 1 are transmitted to the spectrometer, enabling a measurement of the coincidence spectrum.

Figure~\ref{fig:Spectres}, shows the results of the 844~nm spectrum measurements and 775~nm coincidence spectra in three situations: unfiltered, transmitted by the FBG, and reflected by the FBG and passing through the circulator. The results show that the filtering is effectively shaping the 775~nm post-selected spectrum, as expected from tight energy correlations imposed by the narrowband pump. In Figure~\ref{fig:Spectres}~(c), the final 775~nm spectrum is compared to the PPLN acceptance bandwidth, showing that effective effeciency is now only reduced by a factor of 0.96. We therefore expect this filtering to yield an improvement of 1.57 compared to the unfiltered case. 
 
\subsection{Comparison of photon triplet rates}
To demonstrate the improvement in triplet generation rate, we use the experimental setup shown in Fig.~\ref{fig:Montage-triplets} to investigate C-SPDC under filtered and unfiltered conditions. We therefore perform measurements with and without the FBG and the circulator. To compare filtered and unfiltered configurations at the same heralding rate, the detected 844-nm photon rate was maintained at $5\times10^5~\mathrm{s}^{-1}$. This required a first-stage pump power of 1~mW without filtering and 10~mW with filtering. In corresponding 10-minute measurements, the 1530-nm and 1570-nm detectors recorded 71 and 730 twofold coincidences in the unfiltered and filtered configurations, respectively. These twofold coincidences measure photon pairs produced in the second stage and do not include detection of the 844-nm herald photon. The herald filter therefore allowed the first-stage pump power to be increased while maintaining the same heralding rate, resulting in a higher second-stage pair rate. 

\begin{table}[ht]
\centering
\caption{Raw and background-corrected triplet generation rates with and without spectral filtering. Error bars represent one standard deviation assuming Poissonian counting statistics.}
\label{tab:rates}
\begin{tabular}{l c c}
\hline
Condition & Triplets per hour  & Triplets per hour (corrected) \\
\hline
Without spectral filtering & 165 $\pm$ 2 & 165 $\pm$ 3 \\ 
With spectral filtering    & 275 $\pm$ 3 &  270 $\pm$ 3 \\
\hline
\end{tabular}
\end{table}

The final measurement was conducted over 72 hours, using both the filtered and unfiltered configurations in alternating 12 h periods, for a total of 36 hours of measurements for each. Table~\ref{tab:rates} presents the results of the triplet measurements for both cases.
To calculate the rate enhancement, we first correct for accidental coincidences, as these are higher in the filtered case: 4.7/h compared to 0.33/h in the unfiltered case.
We find an increase in the triplet rate by a factor of $1.64 \pm 0.03$, which is close to the theoretically predicted value of 1.57. Note that the experimental error bars account only for Poissonian counting statistics; additional uncertainties, such as system drifts during the measurement period, are not included and may contribute to the observed discrepancy. 
While this increase in rates is relatively modest, this is expected because the PPKTP crystal was initially chosen because of its high spectral compatibility with the PPLN waveguide used in the second stage.
The full benefit of this approach will come from using sources with higher brightness but larger bandwidth as the first stage, which would previously have been disadvantageous due to spectral incompatibility.

\section{Discussion and Conclusion}
These results demonstrate that spectral filtering in the heralding arm can significantly improve the effective efficiency of cascaded spontaneous parametric downconversion.
They highlight that the relevant figure of merit in cascaded SPDC is not simply the total photon flux generated in the first stage, but rather the fraction of heralded photons that are spectrally compatible with the second conversion.
Photons that fall outside the acceptance bandwidth of the second stage contribute to detector saturation without increasing the useful triplet rate.
Coincidence-based filtering therefore acts to redistribute the available photon flux into a spectrally useful subset, improving the conversion efficiency per detected herald rather than the intrinsic brightness of the source.
This improvement comes at the expense of increased pump power in the first stage, as filtering reduces the number of transmitted heralding photons.
In most experiments, this trade-off is favorable because the system is more often limited by detector saturation than by available pump power. 
In regimes where pump power is constrained, however, this approach ensures that the only figures of merit impacting the generated triplet rate are the spectral brightness of the first source and the conversion probability of the second source; spectral compatibility can instead be addressed through appropriate filter selection, independently of detector saturation. 

This is the main implication of our approach: by relaxing the requirement for precise spectral matching between the two stages of a cascaded source, it opens up a wide variety of options for optimizing each stage independently and integrating different types of pair sources.
This will be particularly useful for incorporating either high-brightness sources as a first stage, or narrowband or cavity-enhanced nonlinear devices as the second stage.
In such cases, direct spectral matching through source engineering alone can be challenging, and coincidence-based filtering provides a practical alternative by enabling conditional spectral shaping of the photons. 
The technique could also have applications beyond efficiency enhancement, for example in controlling the spectral properties of photons for multipartite interference experiments.

In conclusion, we have demonstrated that spectral filtering in the coincidence basis can be used to improve detection rates of cascaded spontaneous parametric downconversion. By tailoring the conditional spectrum of the pump photon, we observe and quantify an increase in photon-triplet generation rates when the detection rate in the heralding arm is held constant, in agreement with theoretical predictions based on direct measurement of the created conditional spectra. These results indicate that coincidence-based spectral engineering provides a practical tool for improving the performance and flexibility of cascaded photon sources,   particularly in systems where spectral mismatch or detector limitations constrain overall efficiency.

\section*{Acknowledgments}
  Funding for this work was provided by ResearchNB, the Natural Sciences and Engineering Research Council of Canada (NSERC), the Canada Foundation for Innovation (CFI), and the Canada Research Chairs program.

\bibliographystyle{unsrt}
\bibliography{biblio}

\end{document}